\documentclass[twocolumn,prb,showpacs,floatfix,letterpaper]{revtex4}

\usepackage{graphicx}
\usepackage{amsfonts,color}
\usepackage{amssymb,amsmath}
\usepackage{bm}

\newcommand{\be}{\begin{equation}}
\newcommand{\ee}{\end{equation}}
\newcommand{\bea}{\begin{eqnarray}}
\newcommand{\eea}{\end{eqnarray}}

\begin{document}

\title{Fidelity and superconductivity in two-dimensional $t$-$J$ models}

\author{Marcos Rigol}
\affiliation{Department of Physics, Georgetown University, Washington, DC 
20057, USA}
\author{B. Sriram Shastry}
\affiliation{Department of Physics, University of California, Santa Cruz, 
California 95064, USA}
\author{Stephan Haas}
\affiliation{Department of Physics and Astronomy, University of Southern 
California, Los Angeles, California 90089, USA}

% 75.10.Jm Quantized spin models

\begin{abstract}
We compute the ground-state fidelity and various correlations to gauge the 
competition between different orders in two-dimensional $t$-$J$-type models. 
Using exact numerical diagonalization techniques, these quantities 
are examined for (i) the plain $t$-$J$ and $t$-$t'$-$J$ models, (ii) for 
the $t$-$J$ model perturbed by infinite-range $d$-wave or extended-$s$-wave 
superconductivity inducing terms, and (iii) the $t$-$J$ model, plain 
and with a $d$-wave perturbation, in the presence of non-magnetic 
quenched disorder. Various properties at low hole doping are contrasted 
with those at low electron filling. In the clean case, our results are
consistent with previous work that concluded that the plain $t$-$J$ model 
supports $d$-wave superconductivity. As a consequence of the strong correlations 
present in the low hole doping regime, we find that the magnitude of the 
$d$-wave condensate occupation is small even in the presence of large 
$d$-wave superconductivity inducing terms. In the dirty case, we show the 
robustness of the ground state in the strongly correlated regime 
against disorder.
\end{abstract}

\maketitle

\section{Introduction}
\label{sec:intro}
Understanding the mechanism of high-temperature superconductivity has 
remained a subject of much interest since its experimental discovery in 
the cuprates in 1986.\cite{bednorz86} More recently, this subject has 
received renewed attention following the emergence of the first iron 
based (pnictide) high-temperature superconductor.\cite{kamihara06} It is 
generally believed that high-temperature superconductivity has its roots 
in the interplay of strong correlations and reduced 
dimensionality.\cite{dagotto94,imada98,lee06} However, a full theoretical 
understanding of this phenomenon has proven challenging, and consensus 
regarding its microscopic origin has not yet been 
reached.\cite{norman03,anderson04,anderson07,scalapino06} 

A further complication arises from experimental findings that the doped 
cuprates are highly inhomogeneous.\cite{fischer07} This feature has been 
the subject of numerous recent experimental studies using local probes 
such as scanning tunneling spectroscopy 
(STS).\cite{gomes07,niestemski07,pasupathy08,ma08,kohsaka08,slezak08}  
Theoretical studies of correlations and disorder in superconducting 
lattice models have, in general, either focused on $d$-wave BCS 
phenomenology in the presence of impurities\cite{lee93,sun95,balatsky95} 
or on microscopic disordered $t$-$J$ and Hubbard type models, sometimes 
with the addition of short-ranged terms that favor superconductivity. 
One basic result of BCS phenomenology is that non-magnetic impurities 
suppress superconductivity more strongly in nodal systems, i.e., 
$d$-wave superconductors, than in conventional $s$-wave 
superconductors.\cite{lee93,sun95,balatsky95,anderson59} 
This raises the question why superconductivity in the high-temperature 
cuprates appears to be rather resistant to impurity disorder, although 
they have a $d$-wave superconducting order parameter. A large part of the 
answer presumably involves the short coherence length, which is of the 
order of a few lattice constants in these systems, as opposed to the 
enormous values attained in conventional superconductors. Another problem 
within the standard BCS phenomenology appears to be that the only way 
correlations enter is through superconducting pairing channels, neglecting 
potentially important effects due to the presence of fluctuations towards 
other competing instabilities. The robustness of high-temperature 
superconductivity and its density of states against disorder has been 
recently studied in the framework of Hubbard and $t$-$J$ 
models using Bogoliubov--deGennes\cite{gard08,andersen08} and Gutzwiller 
mean-field theories,\cite{gard08} and exact diagonalization.\cite{rigol09}

In a spirit similar to previous studies,\cite{dagotto94,rigol09} in this work 
we use numerical diagonalization of finite clusters to examine the 
effects of doping a strongly correlated Mott insulator within the $t$-$J$ 
model. The $t$-$J$ model can be justified microscopically either by a 
large $U/t$ expansion of the one-band Hubbard model\cite{dagotto94,lee06} or 
by a reduction of the three-band copper oxide model to an effective 
single-band model.\cite{zhang88,shastry89} The latter approach provides 
greater freedom for the allowed parameter ratio of $J/|t|$. Following 
Ref.\ \onlinecite{rigol09}, we also consider the $t$-$J$ model with the 
addition of an infinite-range superconducting term. This term is tunable, 
and structured to induce either $d$-wave or extended-$s$-wave  
superconductivity. Furthermore, we analyze the effects of quenched 
disorder in the ground state of these systems.

In this work, we focus on three key observables that provide unique 
insights into the properties of the $t$-$J$ model. The first of these 
observables is the ground-state fidelity metric $g$, defined below 
in Eq.\ (\ref{eq:metric}). This quantity is related to the rate of
change of the overlap between the ground states of two Hamiltonians 
induced by a small change of a control parameter. The ground-state 
fidelity, originally studied in the context of quantum information 
theory, has been shown to be a sensitive indicator of changes in 
the ground state of many-body systems, as they occur in quantum phase 
transitions.\cite{zanardi06,cozzini07,you07,buonsante07,chen07,venuti07,venuti08,garnerone09,gu09}
The other two observables of interest we will study are the $d$-wave and 
extended-$s$-wave superconductivity condensate occupations. They will be 
defined carefully in the next section.

A few words motivating this project and the tools used are appropriate 
at this point. By changing control parameters in a quantum many-body 
system, one may encounter first-order or continuous quantum phase 
transition. In finite systems, first-order transitions are easy to 
monitor, using the ground-state energy and density matrices. We 
have studied pair-pair (with $d$-wave and extended-$s$-wave symmetries) 
and density-density/spin-spin correlations. The former indicate 
superconducting instabilities and the latter indicate charge or spin 
orderings. Continuous transitions are more subtle. Due to the 
Wigner--von Neumann non crossing rule, different states of the same 
symmetry approach each other near a transition but do not cross; thereby 
leading to energy gaps. These energy gaps ultimately close for very large 
systems at quantum phase transitions. In recent work, the fidelity 
[Eq.\ (\ref{fidelity})], and especially the associated fidelity metric 
[Eq.\ (\ref{eq:metric})], have been shown to be sensitive indicators of continuous  
transitions,\cite{zanardi06,cozzini07,you07,buonsante07,chen07,venuti07,venuti08,garnerone09} 
and we will see below that these also track level crossings quite well, as 
in general $g$ exhibits a jump in those cases.

The paper is organized as follows. First, the $t$-$J$ model and the
observables of interest are described in Sec.\ \ref{sec:pre}. Sec.\ 
\ref{sec:hom} describes our study of the plain $t$-$J$ and $t$-$t'$-$J$ 
models when tuning the ratio $J/t$. In Sec.\ \ref{sec:sup}, we examine the 
$t$-$J$ model with the addition of $d$-wave and extended-$s$-wave 
superconductivity inducing terms. The effects of quenched disorder in 
the fidelity and superconductivity order parameters are analyzed in Sec.\ 
\ref{sec:dis}. In Sec.\ \ref{sec:sum}, we conclude with a summary of our 
findings. 

\section{Preliminaries}
\label{sec:pre}

\subsection{Model Hamiltonians}

The Hamiltonian for the plain $t$-$J$ model can be written as
\begin{eqnarray}
\hat{{\cal H}}_{tJ}&=&-t\sum_{\langle i,j\rangle,s} 
\hat{P}\left(\hat{c}^\dagger_{is}\hat{c}^{}_{js} 
+ \mathrm{H.c.} \right)\hat{P} \nonumber \\
&&+J \sum_{\langle i,j\rangle} \hat{P}
\left(\hat{{\bf S}}_i\cdot\hat{{\bf S}}_j-\frac{1}{4}\hat{n}_i 
\hat{n}_j \right)\hat{P},
\label{eq:tJ}
\end{eqnarray}
where $\hat{c}^\dagger_{is}$ and $\hat{c}^{}_{is}$ are the creation 
and annihilation operators for an electron with spin $s= \uparrow, 
\downarrow$ on a site $i$,  
$\hat{n}_i=\sum_{s}\hat{c}^\dagger_{is}\hat{c}^{}_{is}$ is the density 
operator, $\hat{P}$ is a projection operator to ensure that there are no 
doubly occupied sites, i.e., it is assumed that the local Coulomb 
repulsion is very large such that two electrons (with antiparallel spin) 
cannot be on the same lattice site, and 
\begin{equation}
\hat{{\bf S}}_i=\frac{1}{2}\sum_{ss'} 
\hat{c}^\dagger_{is}{\bf\sigma}_{ss'} \hat{c}^{}_{is'}
\end{equation}
is the local spin operator (${\bf \sigma}$ are the spin-1/2 Pauli 
matrices). The sums $\langle i,j\rangle$ in Eq.\ (\ref{eq:tJ}) run over 
nearest-neighbor sites.

It has been discussed is several works that longer-range hoppings are 
needed to reproduce the Fermi surface and electron-hole asymmetries for 
different cuprate  
superconductors.\cite{gooding94,nazarenko95,bansil05,belinicher96,lema97,kim98,martins01}
Here we will briefly discuss the effect of the next-nearest-neighbor 
hopping
\begin{equation}
\hat{{\cal H}}_{t'}=-t'\sum_{\langle\langle i,j\rangle\rangle,s} 
\hat{P}\left(\hat{c}^\dagger_{is}\hat{c}^{}_{js} 
+ \mathrm{H.c.} \right)\hat{P},
\label{eq:t'}
\end{equation}
where now the sum $\langle\langle i,j\rangle\rangle$ runs over 
next-nearest-neighbor sites.

Within a number of mean-field theories,\cite{baskaran87,kotliar88,wen96} 
the phase diagram of the doped $t$-$J$ model exhibits $d$-wave 
superconductivity. However, it is not clear how quantum fluctuations, 
which are expected to be significant in two dimensions, would affect such 
a state. As suggested by numerical
diagonalization\cite{dagotto90,dagotto92,dagotto93} and quantum 
Monte-Carlo studies,\cite{hellberg95,hellberg97} competing states with 
other broken symmetries are expected to occur in the proximity of the 
superconducting phase. In particular, there has been a debate within the 
literature whether at low hole doping, phase separation in the 
two-dimensional case occurs already at infinitesimal $J/t>0$ or only 
beyond a finite threshold value,\cite{emery90,hellberg95,hellberg97,sorella98,lugas06} 
preceded by $d$-wave superconductivity.\cite{dagotto90,dagotto92,dagotto93}
It thus appears that the exact phase diagram of the $t$-$J$ model still remains 
to be settled. Here, we address some of these issues using the fidelity 
as an indicator. Meanwhile, in order to precipitate a superconducting 
ground state in the presence of strong correlations, we add attractive 
terms of the form
\begin{equation}
\hat{{\cal H}}_d = - \frac{\lambda_d}{L} \sum_{i,j=1}^L 
\hat{P}\;\hat{\cal D}^\dagger_i \, 
\hat{\cal D}_j\; \hat{P},
\label{eq:dwave}
\end{equation}
which favors a $d$-wave superconducting pairing, and 
\begin{equation}
\hat{{\cal H}}_s = - \frac{\lambda_s}{L} \sum_{i,j=1}^L \hat{P}\; \hat{\cal S}^\dagger_i \, 
\hat{\cal S}_j\;\hat{P},
\label{eq:swave}
\end{equation}
which favors an extended-$s$-wave superconducting pairing. 

In Eqs.\ (\ref{eq:dwave}) and (\ref{eq:swave}),
$\hat{\cal D}_i=(\hat{\Delta}_{i,i+\hat{\textrm{x}}}-\hat{\Delta}_{i,i+\hat{\textrm{y}}})$,
$\hat{\cal S}_i=(\hat{\Delta}_{i,i+\hat{\textrm{x}}}+\hat{\Delta}_{i,i+\hat{\textrm{y}}})$ 
and $\hat{\Delta}_{ij}=\hat{c}_{i\uparrow}\hat{c}_{j\downarrow}+
\hat{c}_{j\uparrow}\hat{c}_{i\downarrow}$. $\lambda_d$ and $\lambda_s$ 
denote the strengths of the $d$-wave and extended-$s$-wave 
superconductivity inducing terms, respectively, and  $L$ is the number of 
lattice sites. Equations (\ref{eq:dwave}) and (\ref{eq:swave}) have 
infinite-range terms of the type Bardeen, Cooper, and Schrieffer 
considered in their reduced Hamiltonian,\cite{bcs57} while building in 
the $d$-wave and extended-$s$-wave symmetries in the superconductivity 
order parameter. Within mean-field theory, Eq.\ (\ref{eq:tJ}) with the 
addition of Eq.\ (\ref{eq:dwave}) leads to the same $d$-wave ground state 
as found from the $t$-$J$ model.\cite{baskaran87,kotliar88} 

Finally, we also consider the effects of quenched random disorder of the 
form
\begin{equation}
\hat{{\cal H}}_{random}   =  \sum_{i} \varepsilon_i \hat{n}_i ,
\label{eq:disorder}
\end{equation}
where the $\varepsilon_i$'s are taken randomly from a uniform distribution 
between [$-\Gamma,\Gamma$]. The full Hamiltonians given by Eqs.\ 
(\ref{eq:tJ})--(\ref{eq:disorder}) thus describe inhomogeneous strongly 
correlated superconductors. In this study, we use numerical 
diagonalization of clusters with 18 and 20 sites and periodic boundary 
conditions. The cluster geometries can be found in Ref.\ 
\onlinecite{dagotto94}. The dimension of the largest Hilbert space 
diagonalized is of the order of $10^8$. 

\subsection{Observables}

The first observable of interest is related to the fidelity $F$, which is 
defined as follows. Assume a general Hamiltonian of the form
\[
 \hat{\cal H}(\lambda) = \hat{\cal H}_0 + \lambda\, \hat{\cal H}_1,
\]
where $\hat{{\cal H}}_1$ is taken to be the driving term. In the next 
sections, we will consider $\hat{{\cal H}}_1$ to be either the Heisenberg 
interaction term in Eq.\ (\ref{eq:tJ}), the superconducting terms given 
by Eqs.\ (\ref{eq:dwave}) and (\ref{eq:swave}), or the disorder term in 
Eq.\ (\ref{eq:disorder}). Let $\vert\Psi_0(\lambda)\rangle$ be the 
(normalized) ground state of $\hat{{\cal H}}(\lambda)$ and 
$\vert\Psi_0(\lambda+\delta\lambda)\rangle$ be the 
(normalized) ground state of $\hat{{\cal H}}(\lambda+\delta\lambda)$. The 
fidelity is then defined as the overlap between 
$\vert\Psi_0(\lambda)\rangle$ and 
$\vert\Psi_0(\lambda+\delta\lambda)\rangle$, i.e., 
\be
F(\lambda,\delta\lambda) = \vert \langle\Psi_0(\lambda) \vert 
\Psi_0(\lambda+\delta\lambda)\rangle\vert.
\label{fidelity}
\ee

If the ground state is nondegenerate, and if $\delta\lambda$ is 
sufficiently small, one can compute  
$\vert\Psi_0(\lambda+\delta\lambda)\rangle$ up to second order 
in perturbation theory. The only two terms of the (normalized) second 
order expansion that have a nonvanishing overlap (n.o.) with 
$\vert\Psi_0(\lambda)\rangle$ are 
\bea
&&\vert\Psi_0(\lambda+\delta\lambda)\rangle_{\textrm{n.o.}} =  
\vert\Psi_0(\lambda)\rangle \times \\ && \qquad\qquad\qquad\ \ \ \left( 
1-\dfrac{\delta\lambda^2}{2} \sum_{\alpha \neq 0} 
\dfrac{\vert \langle\Psi_\alpha(\lambda) \vert \hat{\cal H}_1 \vert 
\Psi_0(\lambda)\rangle\vert^2}
{\left[ E_0(\lambda)-E_\alpha(\lambda)\right] ^2}\right)\nonumber
\eea
where $\vert\Psi_\alpha(\lambda)\rangle$ are the eigenstates of the 
Hamiltonian with eigenenergies $E_\alpha(\lambda)$, i.e., $\hat{{\cal 
H}}(\lambda) \vert\Psi_\alpha(\lambda)\rangle=E_\alpha(\lambda) 
\vert\Psi_\alpha(\lambda)\rangle$.

This means that up to the lowest order in $\delta\lambda$, one can write 
the fidelity in the form
\be
F(\lambda,\delta\lambda) = 1-\dfrac{\delta\lambda^2}{2}\sum_{\alpha \neq 
0} 
\dfrac{\vert \langle\Psi_\alpha(\lambda) \vert \hat{\cal H}_1 \vert 
\Psi_0(\lambda)\rangle\vert^2}
{\left[ E_0(\lambda)-E_\alpha(\lambda)\right] ^2}.
\ee
Since the sum on the rhs is in most cases an extensive quantity (see, e.g., 
Refs.\ \onlinecite{you07}, \onlinecite{venuti07}, and \onlinecite{venuti08}; 
for counterexamples see, e.g., Ref.\ \onlinecite{gu09}), 
one can define the fidelity metric as
\begin{eqnarray}
 g(\lambda,\delta\lambda) &\equiv& \dfrac{2}{L} 
\dfrac{1-F(\lambda,\delta\lambda)}{\delta\lambda^2}  \nonumber \\
\lim_{\delta \lambda \rightarrow 0}g(\lambda,\delta\lambda) &=&  \dfrac{1}{L}\sum_{\alpha \neq 0} 
\dfrac{\vert \langle\Psi_\alpha(\lambda) \vert \hat{\cal H}_1 \vert 
\Psi_0(\lambda)\rangle\vert^2}
{\left[ E_0(\lambda)-E_\alpha(\lambda)\right] ^2}. 
\;\; \label{eq:metric}
\end{eqnarray}
In the following, we refer to $ \lim_{\delta \lambda \rightarrow 0} 
g(\lambda, \delta \lambda)$  as $g(\lambda)$ or simply as $g$. From its 
definition, the fidelity metric $g$ is dimensionless, positive and 
(in most cases) intensive, i.e., of $O(1)$. This is one of the main quantities 
that we will examine in the following sections. $F(\lambda,\delta\lambda)$ 
will be computed using Lanczos diagonalization, choosing a value of 
$\delta\lambda$ that is sufficiently small so that it does not affect the 
result of the ratio in Eq.\ (\ref{eq:metric}), i.e., giving effectively 
the value in the $\lim_{\delta \lambda \rightarrow 0} g(\lambda, 
\delta \lambda)$. The above is of course true provided one does not 
encounter a level crossing. At a crossing, we compute $g$ on either side
of its jump by the above limiting process.

The other two quantities of interest are the $d$-wave and 
extended-$s$-wave superconductivity condensate occupations. Given the 
$d$-wave pair density matrix
\be
P_{ij}^d=\langle\Psi_0 \vert \hat{P}\;\hat{\cal D}^\dagger_i \, \hat{\cal 
D}_j\; \hat{P} \vert \Psi_0\rangle,
\ee
and the extended-$s$-wave pair density matrix
\be
P_{ij}^s=\langle\Psi_0 \vert \hat{P}\;\hat{\cal S}^\dagger_i \, \hat{\cal 
S}_j\; \hat{P} \vert \Psi_0\rangle,
\ee
the $d$-wave ($\Lambda_1^d$) and extended-$s$-wave ($\Lambda_1^s$) 
condensate occupations are defined as the largest eigenvalues of 
$P_{ij}^d$ and $P_{ij}^s$.\cite{penrose56} The corresponding eigenvectors 
of the density matrices are known as the ``natural orbitals,'' and those 
with the largest eigenvalues are referred to as the ``lowest natural 
orbitals.\cite{lowdin55}'' If a condensate of pairs with a particular 
symmetry occurs in the system, the corresponding condensate occupation 
will scale linearly with the total number of fermions, as the system size 
$L$ is increased while keeping the density constant.\cite{penrose56} This 
in turn is equivalent to stating that $P_{ij}^d$ and $P_{ij}^s$ exhibit 
off-diagonal long-range order.\cite{yang62} Condensation also implies 
that all other eigenvalues are $\Lambda_\alpha^{d,s}\sim 
O(1)$.\cite{leggett01} An advantage of using these definitions is that 
they are valid independently of whether the system is translationally 
invariant or not, i.e., they work the same in clean systems and in the 
presence of disorder. In the particular case of translationally invariant 
systems, the eigenvalues $P_{ij}^d$ and $P_{ij}^s$ are occupations in 
momentum space.

Since we will be dealing here with systems with different densities and 
finite sizes, in many cases we find it useful to monitor the ratios 
$R^d=\Lambda_1^d/\Lambda_2^d$ and $R^s=\Lambda_1^s/\Lambda_2^s$ between 
the largest eigenvalues ($\Lambda_1^d$, $\Lambda_1^s$) and the second 
largest eigenvalues ($\Lambda_2^d$, $\Lambda_2^s$) of the density 
matrices. These ratios were first introduced in our earlier 
work,\cite{rigol09} and here we briefly reiterate the motivation behind 
this construction. If condensation occurs, i.e., symmetry is broken in the
thermodynamic limit, these are equivalent to studying $\Lambda_1^d$ and 
$\Lambda_1^s$ because the next eigenvalue is small, i.e., 
$\Lambda_2^{d,s}\sim O(1)$. However, computing $R^d$ and $R^s$ has the 
added benefit of eliminating uninteresting normalization effects related 
to the change in the particle density, etc. It also has some advantages 
when trying to understand the effects of changing a Hamiltonian parameter 
for a system with a fixed size, where we find cases with $\Lambda_1$ and 
$R$ behaving differently.

\section{Plain $t$-$J$ model}
\label{sec:hom}

As a first step, in this section we study how the observables of interest 
behave within the plain $t$-$J$ model. We begin with the effect of the 
antiferromagnetic Heisenberg coupling on the ground state of this system. 
Within many mean-field theories, finite values of $J$ favor 
superconducting ground states close to  
half-filling.\cite{baskaran87,kotliar88,wen96} The energy scale is set by 
the hopping parameter $t=1$. We further study the ground state of the 
$t$-$J$ model for values of $J$ between 0 and 1. While this rather large 
range is not achievable within the large $U/t$ expansion of the one-band 
Hubbard model (where $J\sim t^2/U$), it should rather be regarded as a 
``Gedanken range,'' intended for the purpose of studying the 
possible phases that emerge, including the widely discussed  possibility 
of phase separation.\cite{emery90,sorella98,dagotto93}

We begin by computing the fidelity metric in Eq.\ (\ref{eq:metric}), 
considering the Heisenberg coupling in Eq.\ (\ref{eq:tJ}) to be 
the tuning parameter in the Hamiltonian, so that $J$ plays the role of 
$\lambda$. For our computations, we find that taking $\delta J=10^{-5}$ 
is small enough to provide results for $g$ which are independent of $\delta J$.

\begin{figure}[!ht]
\begin{center}
  \includegraphics[width=0.41\textwidth,angle=0]{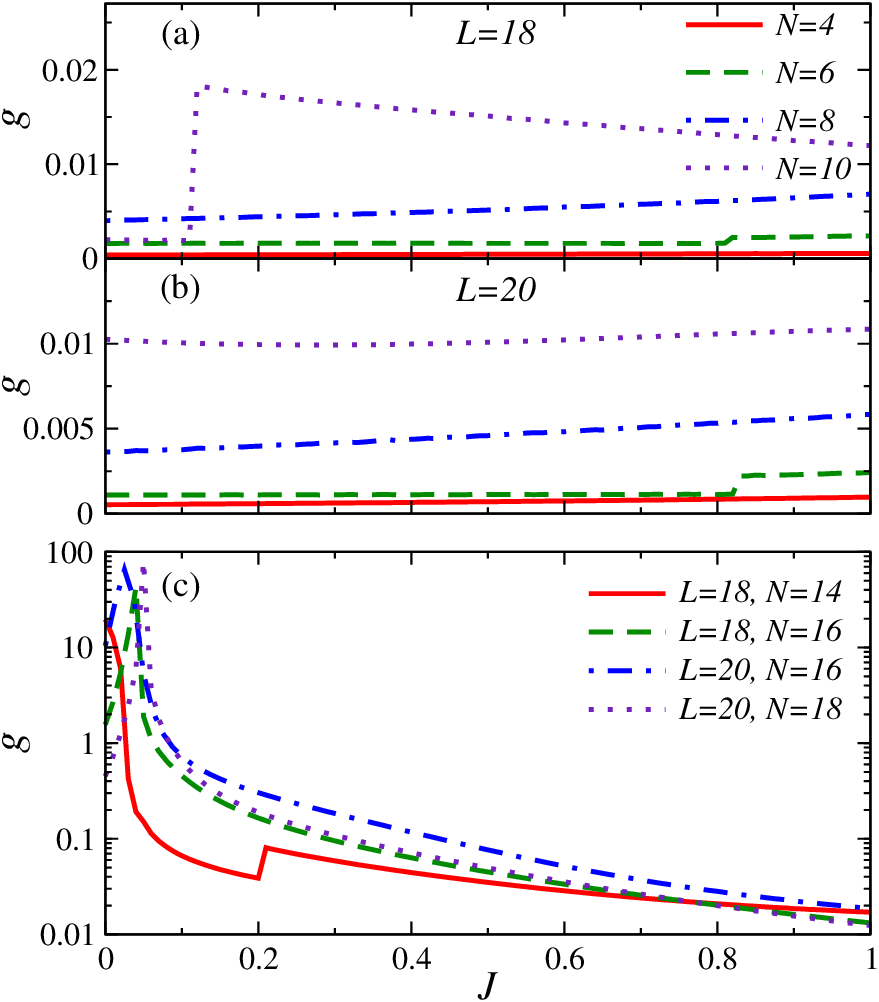}
\end{center}
\vspace{-0.6cm}
\caption{\label{fig:FidelityvsJ} (Color online) Fidelity metric of the 
$t$-$J$ model as a function of the antiferromagnetic coupling $J$. (a) 
Cluster with 18 sites and fillings $N=4,6,8,10$. (b) Cluster with 20 
sites and fillings $N=4,6,8,10$ [the same parameter sets as in (a)]. (c) 
Clusters with 18 and 20 sites and two and four holes. Notice that for all 
our results in this paper we consider $N=N_\uparrow+N_\downarrow$, 
where $N_\uparrow=\langle \hat{c}^\dagger_{\uparrow}\hat{c}^{}_{\uparrow}\rangle$,
$N_\downarrow=\langle \hat{c}^\dagger_{\downarrow}\hat{c}^{}_{\downarrow}\rangle$, and 
$N_\uparrow=N_\downarrow$.}
\end{figure}

In Fig.\ \ref{fig:FidelityvsJ}, we compare the dependence of the fidelity 
metric on $J$ for low electron filling [Figs.\  
\ref{fig:FidelityvsJ}(a) and \ref{fig:FidelityvsJ}(b)] and low hole filling 
[Fig.\ \ref{fig:FidelityvsJ}(c)]. The contrast between the two cases is 
very clear. For low electron filling the fidelity changes very little 
between $J=0$ and $1$, but we do observe some level crossings, signaled by 
relatively small jumps in the fidelity metric. These have their origin in 
the peculiarities of particular cluster shapes, as exemplified by the 
fact that for the same particle number they are found in one of 
the clusters and not in the other (see, e.g., the case of ten particles), 
and therefore they are expected to disappear in the thermodynamic limit. 

%%%%%%%%%%%%%%  FIGURE  %%%%%%%%%%%%%%%%%%%%%%%%%%%%%%%%%%%%%%%%%%%%%%
%%%%%%%%%%%%%%%%%%%%%%%%%%%%%%%%%%%%%%%%%%%%%%%%%%%%%%%%%%%%%%%%%%%%%%%%

For low hole doping, on the other hand, see Fig.\ \ref{fig:FidelityvsJ}(c), 
the fidelity metric exhibits a large response for small $J \lesssim 0.05 $, 
suggestive of a continuous phase transition in that regime. The cluster sizes 
accessible to numerical diagonalization are too small to identify a critical 
point, if any, but we find that a qualitatively similar behavior occurs for 
two and four holes in both clusters with 18 and 20 sites. 

%%%%%%%%%%%%%%  FIGURE  %%%%%%%%%%%%%%%%%%%%%%%%%%%%%%%%%%%%%%%%%%%%%%
\begin{figure}[!b]
\begin{center}
  \includegraphics[width=0.48\textwidth,angle=0]{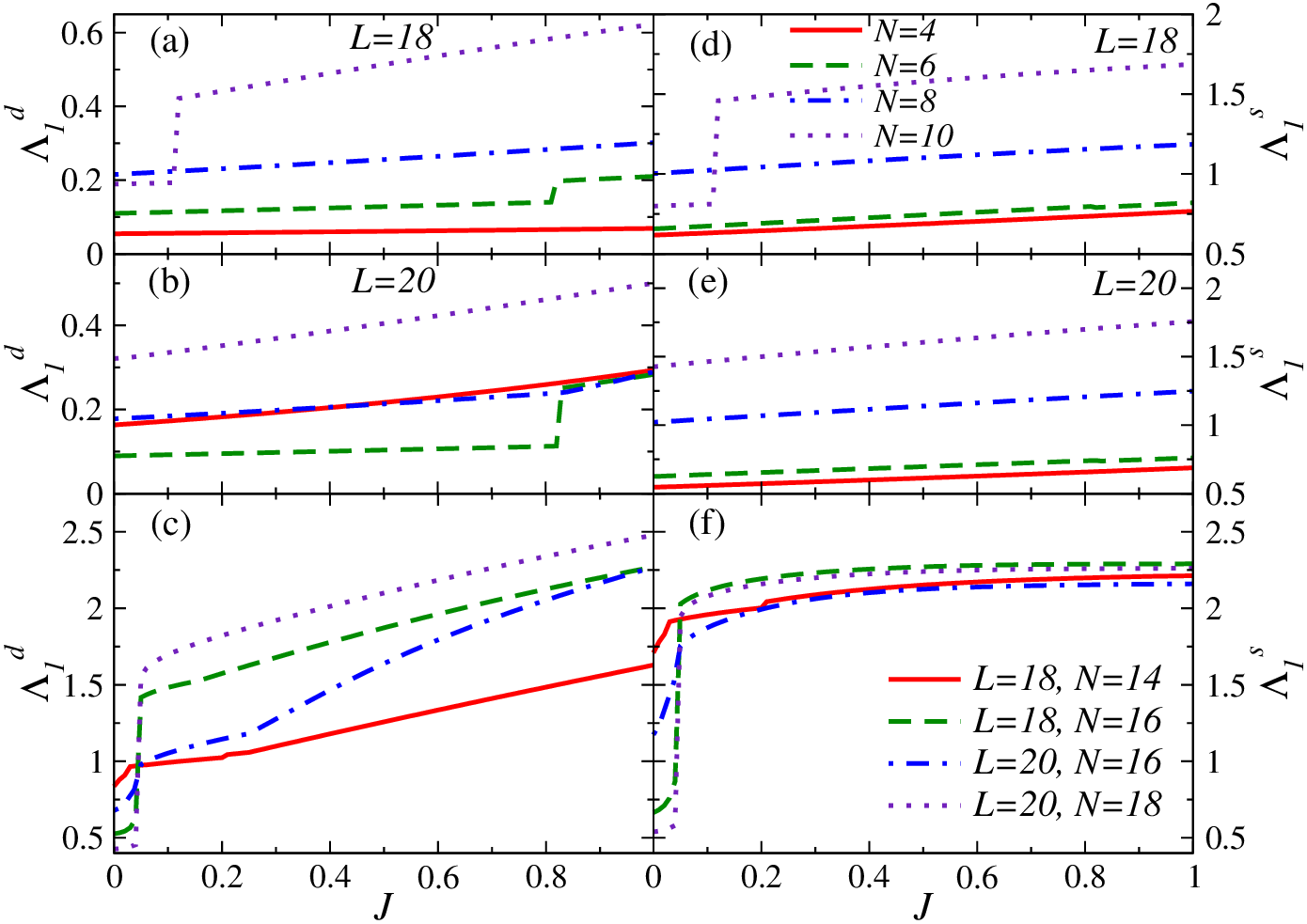}
\end{center}
\vspace{-0.6cm}
\caption{\label{fig:NaturalOrbital01vsJ} (Color online) Occupation of the 
lowest natural orbital, of the $d$-wave (left panels) and 
extended-$s$-wave (right panels) density matrices, as a function of $J$. 
[(a) and (d)] Cluster with 18 sites and 
fillings $N=4,6,8,10$. [(b) and (e)] Cluster with 20 sites and fillings 
$N=4,6,8,10$. Notice that the parameter sets for (a)--(e) are all the same and 
given in (d). [(c) and (f)] Clusters with 18 and 20 sites and two and four 
holes. The parameter sets in (c) and (f) are also the same.}
\end{figure}
%%%%%%%%%%%%%%%%%%%%%%%%%%%%%%%%%%%%%%%%%%%%%%%%%%%%%%%%%%%%%%%%%%%%%%%%

Next consider the occupation of the $d$-wave and extended-$s$-wave lowest 
natural orbitals as $J$ is tuned in the system.  Results for these 
quantities are shown in Fig.\ \ref{fig:NaturalOrbital01vsJ}. As observed 
here, changes in the fidelity metric in Fig.\ \ref{fig:FidelityvsJ} are 
accompanied by changes in the lowest natural orbital occupations in Fig.\ 
\ref{fig:NaturalOrbital01vsJ}. Interestingly, for each given cluster 
shape and filling fraction, the behavior of $\Lambda_1^d (J)$ and 
$\Lambda_1^s(J)$ is qualitatively similar. We do find, however, a marked 
difference between the results obtained for $\Lambda_1^d$ and 
$\Lambda_1^s$  between low electron filling, and low 
hole doping. For low electron filling [Figs.\ 
\ref{fig:NaturalOrbital01vsJ}(a), \ref{fig:NaturalOrbital01vsJ}(b),
\ref{fig:NaturalOrbital01vsJ}(d), and \ref{fig:NaturalOrbital01vsJ}(e)], 
the lowest natural orbital occupation only changes appreciably when a 
level crossing is observed in the fidelity (Fig.\ \ref{fig:FidelityvsJ}). 
Also, in this case, we find $\Lambda_1^s > \Lambda_1^d$, suggesting 
dominant $s$-wave order parameter that is consistent with the results in 
Refs.\ \onlinecite{dagotto92} and \onlinecite{dagotto93}. In contrast, 
for low hole doping [Figs.\ \ref{fig:NaturalOrbital01vsJ}(c) and 
\ref{fig:NaturalOrbital01vsJ}(f)] with two and four holes in both cluster 
geometries, the occupation of the lowest natural orbitals in all cases 
exhibit a large increase for small values of $J$ ($J \lesssim 0.05$), where 
a sizable response was observed in $g$ in Fig.\ \ref{fig:FidelityvsJ}.

%%%%%%%%%%%%%%  FIGURE  %%%%%%%%%%%%%%%%%%%%%%%%%%%%%%%%%%%%%%%%%%%%%%
\begin{figure}[!h]
\begin{center}
  \includegraphics[width=0.48\textwidth,angle=0]{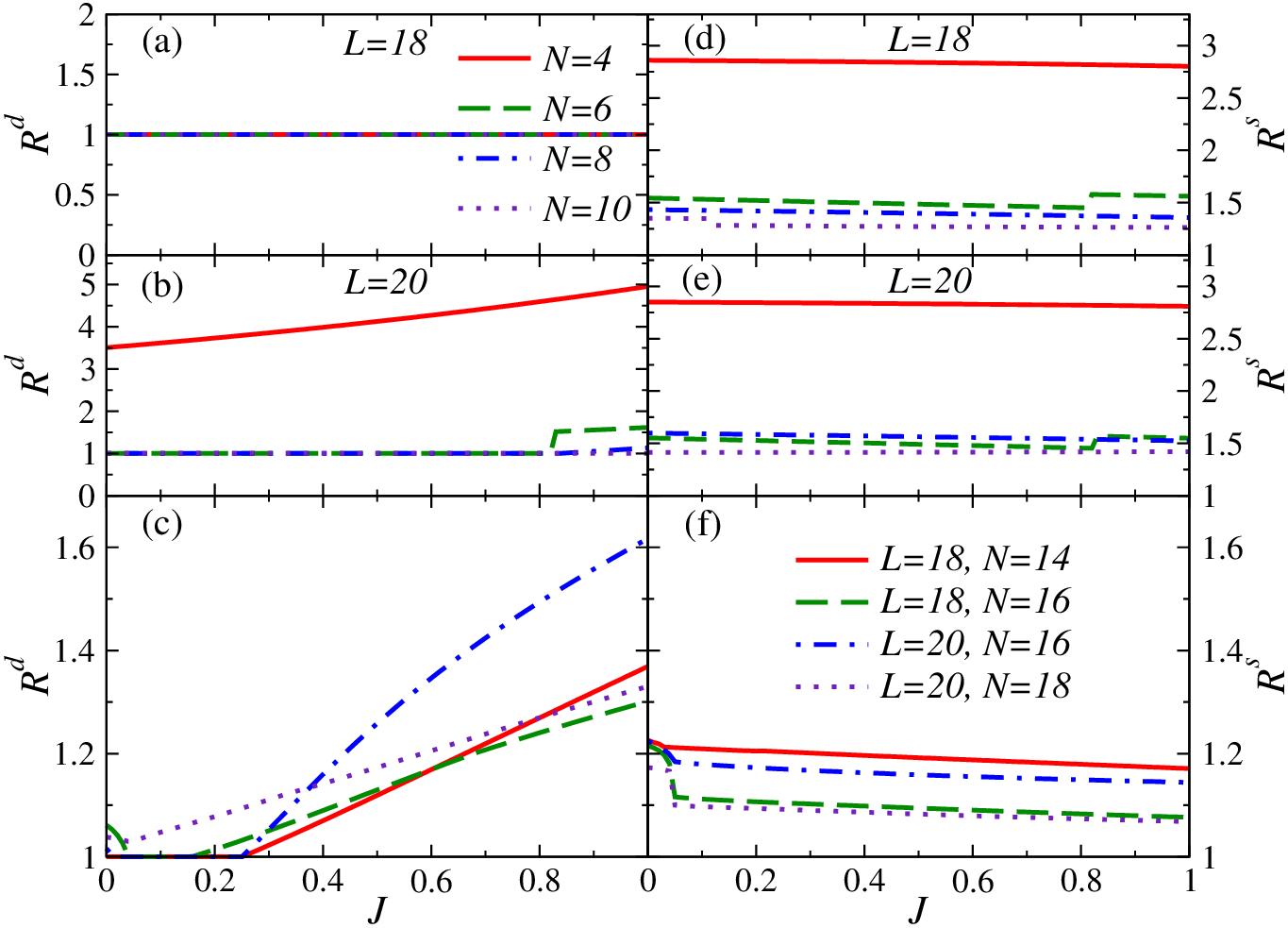}
\end{center}
\vspace{-0.6cm}
\caption{\label{fig:NaturalOrbitalRatiovsJ} (Color online) Ratio between the 
two lowest natural orbitals, of the $d$-wave (left panels) and extended-$s$-wave 
(right panels) density matrices, as a function of $J$. 
[(a) and (d)] Cluster with 18 sites and 
fillings $N=4,6,8,10$. [(b) and (e)] Cluster with 20 sites and fillings 
$N=4,6,8,10$. Notice that the parameter sets for (a)--(e) are all the same and 
given in (a). [(c) and (f)] Clusters with 18 and 20 sites and 
two and four holes. The parameter sets in (c) and (f) are also the same.}
\end{figure}
%%%%%%%%%%%%%%%%%%%%%%%%%%%%%%%%%%%%%%%%%%%%%%%%%%%%%%%%%%%%%%%%%%%%%%%%

It is interesting to observe that the occupation of the lowest natural 
orbitals of both $d$-wave and extended-$s$-wave symmetries in general 
increase as a result of increasing $J$. Since we cannot perform finite 
size scaling by studying larger systems, this does not throw light 
on the competition between these two orderings. In order to resolve
this issue, we have also studied the ratio between the lowest two 
natural orbitals for both superconducting symmetries. The results are 
presented in Fig.\ \ref{fig:NaturalOrbitalRatiovsJ}. They show a quite 
different behavior from the one seen for the lowest natural orbitals in 
Fig.\ \ref{fig:NaturalOrbital01vsJ}. We see from Figs.\ 
\ref{fig:NaturalOrbitalRatiovsJ}(a), \ref{fig:NaturalOrbitalRatiovsJ}(b),
\ref{fig:NaturalOrbitalRatiovsJ}(d), and \ref{fig:NaturalOrbitalRatiovsJ}(e), 
that for low electron filling the ratio $R$ between the two lowest natural 
orbitals in general remains almost unchanged when $J$ takes values between 
0 and 1, for both $d$-wave symmetry and extended-$s$-wave symmetry. For low 
hole doping, Figs.\ \ref{fig:NaturalOrbitalRatiovsJ}(c) 
and \ref{fig:NaturalOrbitalRatiovsJ}(f), the behavior is 
different. $R^d$ increases for both two and four holes in 18 and 20 lattice 
sites, whereas $R^s$ decreases for the same set of parameters. This is a strong 
indication that for larger systems sizes the model may select, if any, a 
$d$-wave superconducting ground state. To further explore this possibility, 
in the next section we perturb the $t$-$J$ model Hamiltonian with $d$-wave 
and extended-$s$-wave superconductivity inducing terms and study its response.

\begin{figure}[!h]
\begin{center}
  \includegraphics[width=0.485\textwidth,angle=0]{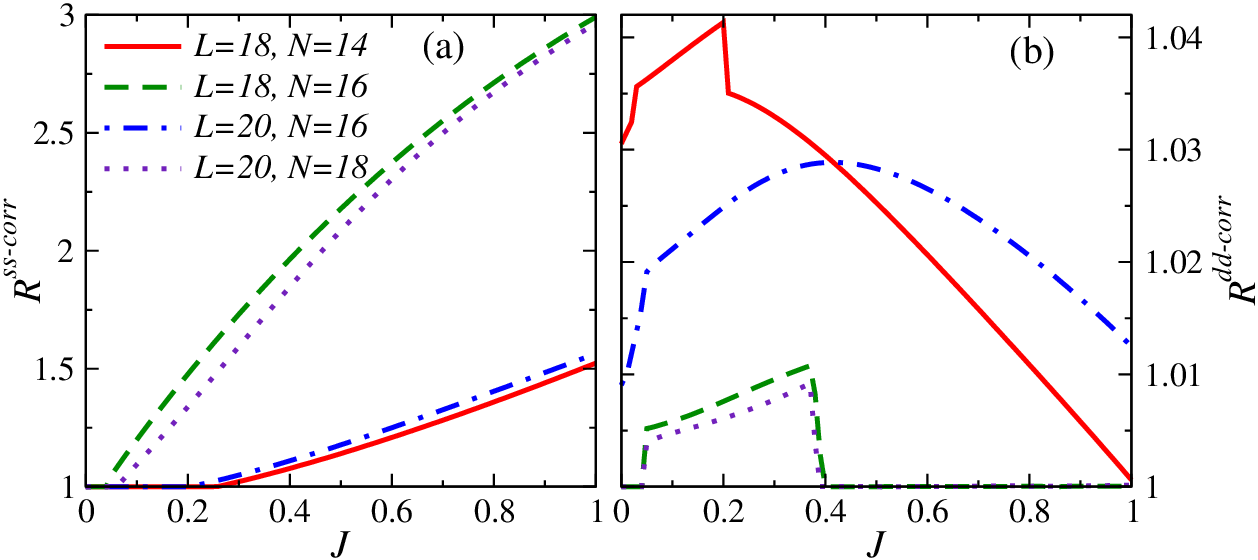}
\end{center}
\vspace{-0.6cm}
\caption{\label{fig:NaturalOrbitalDDRatiovsJ} (Color online) Ratio 
between the two lowest natural orbitals as a function of $J$ in clusters 
with two and four holes with 18 and 20 lattice sites for (a) the 
spin-spin correlation matrix and (b) the density-density correlation 
matrix.}
\end{figure}

We have also studied other correlation functions, such as the spin-spin 
and density-density correlations. The results for the ratio between the 
two lowest natural orbitals of those two matrices are presented in Fig.\ 
\ref{fig:NaturalOrbitalDDRatiovsJ} for two and four holes in 18 and 20 
sites. As expected, spin-spin correlations ($R^{ss\textrm{-}corr}$) 
exhibit a large enhancement corresponding to the onset of  
antiferromagnetic order. $R^{ss\textrm{-}corr}$ reduces dramatically as 
the hole doping increases, in contrast to the behavior of $R^d$ in 
Fig.\ \ref{fig:NaturalOrbitalRatiovsJ}(c), which for large $J$ increases 
as the doping increases (in the underdoped regime). Interestingly, stripes 
are expected to be one of the competing orders at low doping, and they 
should be reflected in the density-density structure factor 
(in $R^{dd\textrm{-}corr}$). In Fig.\ 
\ref{fig:NaturalOrbitalDDRatiovsJ}(b), one can see that 
$R^{dd\textrm{-}corr}$ actually decreases with increasing $J$, i.e., no 
signature of charge-density order can be seen in our clusters.

%%%%%%%%%%%%%%  FIGURE  %%%%%%%%%%%%%%%%%%%%%%%%%%%%%%%%%%%%%%%%%%%%%%
%%%%%%%%%%%%%%%%%%%%%%%%%%%%%%%%%%%%%%%%%%%%%%%%%%%%%%%%%%%%%%%%%%%%%%%%

\subsection{Next-nearest-neighbor ($t'$) term}

As mentioned in Sec.\ \ref{sec:pre}, several authors have discussed the 
importance of introducing long-range hopping in microscopic Hubbard and 
$t$-$J$-like models in order to capture many of the features observed in 
high-temperature 
superconductors.\cite{gooding94,nazarenko95,belinicher96,lema97,kim98,martins01} 
Here, we briefly discuss how a finite next-nearest-neighbor hopping 
affects the results discussed above for the plain $t$-$J$ 
model.

\begin{figure}[!ht]
\begin{center}
  \includegraphics[width=0.485\textwidth,angle=0]{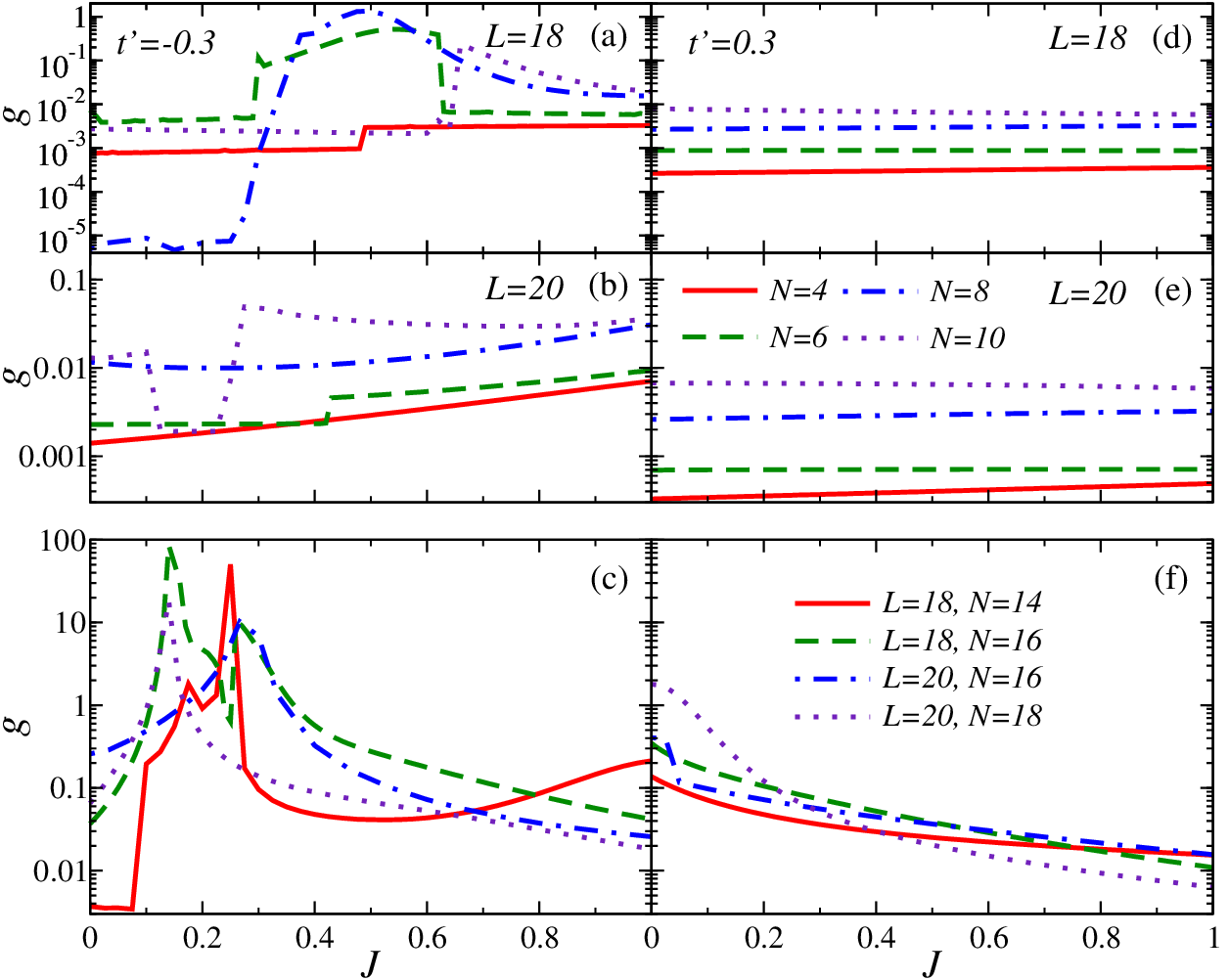}
\end{center}
\vspace{-0.6cm}
\caption{\label{fig:FidelityvsJ_Finitetp} (Color online) Fidelity metric 
of the $t$-$t'$-$J$ model as a function of the antiferromagnetic coupling $J$
for (a)--(c) $t'=-0.3$ and (d)--(f) $t'=0.3$. [(a) and (d)] Cluster with 18 sites 
and fillings $N=4,6,8,10$. [(b) and (e)] Cluster with 20 sites and fillings 
$N=4,6,8,10$ [the legends in (a), (b), and (d) are the same as in (e)]. 
[(c) and (f)] Clusters with 18 and 20 sites and two and four holes (the legends 
are the same in both figures).}
\end{figure}

In Fig.\ \ref{fig:FidelityvsJ_Finitetp}, we show the fidelity metric for 
exactly the same clusters and fillings as in Fig.\ \ref{fig:FidelityvsJ}. 
The results in the presence of a finite and small $t'$ are qualitatively 
similar to those of the plain $t$-$J$ model [Fig.\ \ref{fig:FidelityvsJ}] 
in both the low-electron-filling and low-hole-doping regimes, and for 
negative (left column) and positive (right column) values of $t'$. 
It is worth noticing, however, that in the low doping regime the presence 
of a finite $t'<0$ [Fig.\ \ref{fig:FidelityvsJ_Finitetp}(c)] moves the 
maximum response of the fidelity metric towards larger values of $J$, 
while no such effect is seen for $t'>0$ [Fig.\ \ref{fig:FidelityvsJ_Finitetp}(f)]. 
The former displacement for $t'<0$ is more pronounced for four holes than for 
two holes in both clusters with 18 and 20 sites. This is consistent with previous 
works which have pointed out that in hole-doped systems $t'$ competes with 
$J$ as it suppresses both antiferromagnetic correlations\cite{tohyama94,tohyama04,spanu08} 
and the superconducting $T_c$,\cite{spanu08,khatami08} and that the 
effect of $t'$ becomes stronger with doping (in the low-hole-doping regime).
A positive $t'$, on the other hand, enhances antiferromagnetic correlations,
which is consistent with the results in [Fig.\ \ref{fig:FidelityvsJ_Finitetp}(f)]
where the maximum response of the fidelity is in all cases seen at $J=0$.

%%%%%%%%%%%%%%  FIGURE  %%%%%%%%%%%%%%%%%%%%%%%%%%%%%%%%%%%%%%%%%%%%%%
%%%%%%%%%%%%%%%%%%%%%%%%%%%%%%%%%%%%%%%%%%%%%%%%%%%%%%%%%%%%%%%%%%%%%%%%

The results for the lowest natural orbitals and the ratio between the 
lowest the second lowest natural orbitals are also qualitatively similar 
to those of the plain $t$-$J$ model, and will not be presented here. In 
the remaining of the paper, we will set $t'=0$ and study the effects 
of superconductivity enhancing terms and disorder in the $t$-$J$ model.

\section{Superconductivity Inducing Terms}
\label{sec:sup}

\subsection{$d$-wave term}
\label{sec:supd}

Let us first consider a total Hamiltonian that is the sum of Eq.\ 
(\ref{eq:tJ}) and the $d$-wave-superconductivity inducing term in 
Eq.\ (\ref{eq:dwave}), and study the ground state of this model as a 
function of increasing the parameter $\lambda_d$. In the following, we fix 
the Heisenberg coupling to be $J=0.3$, which is a value commonly used in 
the $t$-$J$ model literature. From the analysis in the previous section, 
we know that, at least for the finite clusters considered here, no 
further qualitative changes occur in the observables of interest for 
larger values of $J$.

Recall from previous work,\cite{rigol09} that the added $d$-wave term 
Eq.\ (\ref{eq:dwave}), being of infinite range, must certainly  
precipitate superconductivity in the $d$-wave channel. This is because 
mean-field theory becomes exact in the thermodynamic limit for an 
infinite-range model of this type. (The same argument, of course, also 
works in the presence of an extended-$s$-wave term.) While this argument 
is true for very large systems, for finite systems one may need a finite 
$\lambda_d \sim O(1)$ to achieve superconductivity.\cite{notes} Therefore, 
we expect that in some cases the fidelity metric should show an enhancement 
as a function of $\lambda_d$ at some characteristic value $\lambda_d^*$. 
A small value of $\lambda_d^*$, consistent with $\lambda_d^*\sim 0$, 
may be taken as an indicator of the incipient order of the $\lambda_d=0$ 
model (the plain $t$-$J$ model).

\begin{figure}[!hb]
\begin{center}
  \includegraphics[width=0.41\textwidth,angle=0]{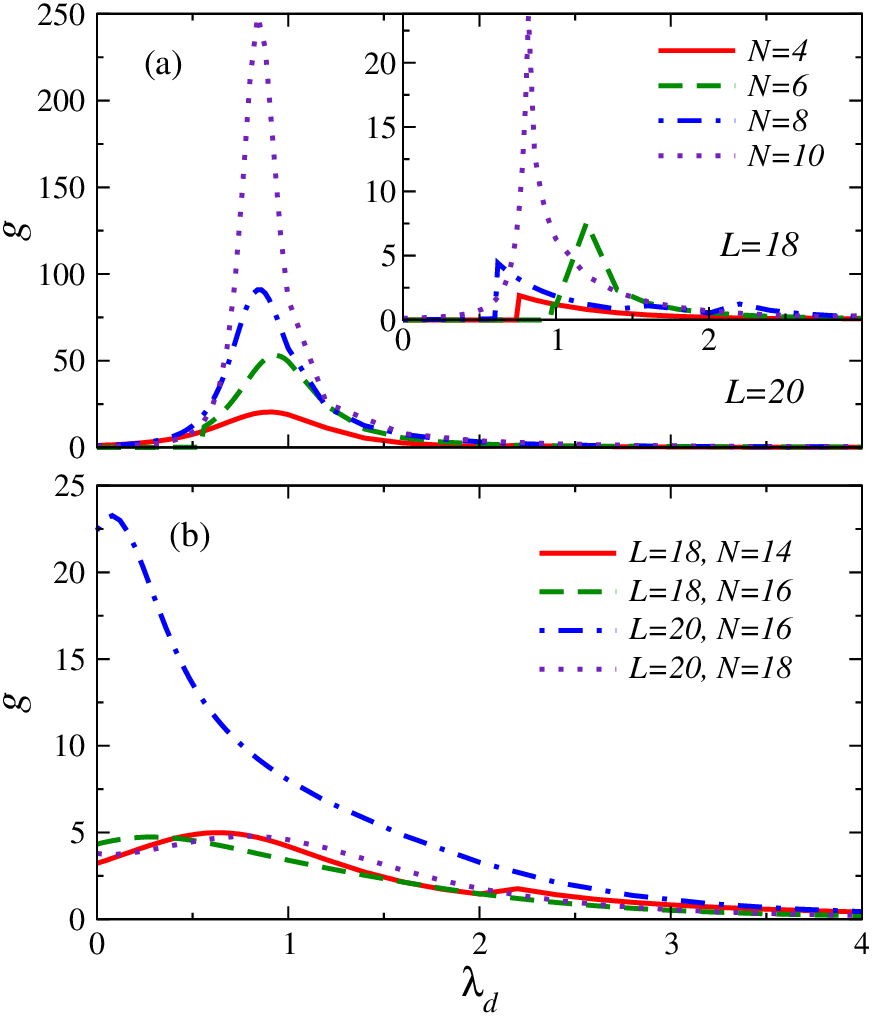}
\end{center}
\vspace{-0.6cm}
\caption{\label{fig:FidelityvsLambda_dwave} (Color online) Fidelity 
metric of the $t$-$J$ model ($J=0.3$) after the addition of a 
$d$-wave-superconductivity inducing term. (a) Cluster with 20 sites and 
fillings $N=4,6,8,10$. (b) Clusters with 18 and 20 sites and two and four 
holes. The inset in (a) depicts the results for the cluster with 18 sites 
and fillings $N=4,6,8,10$.}
\end{figure}

In Fig.\ \ref{fig:FidelityvsLambda_dwave}, we show the fidelity metric as 
a function of the driving parameter $\lambda_d$. In our calculations, we 
have taken $\delta\lambda_d =10^{-5}$, which is sufficiently small to 
ensure results consistent with the limit $\delta\lambda_d\rightarrow 0$.

%%%%%%%%%%%%%%  FIGURE  %%%%%%%%%%%%%%%%%%%%%%%%%%%%%%%%%%%%%%%%%%%%%%
%%%%%%%%%%%%%%%%%%%%%%%%%%%%%%%%%%%%%%%%%%%%%%%%%%%%%%%%%%%%%%%%%%%%%%%%

\begin{figure}[!ht]
\begin{center}
  \includegraphics[width=0.48\textwidth,angle=0]{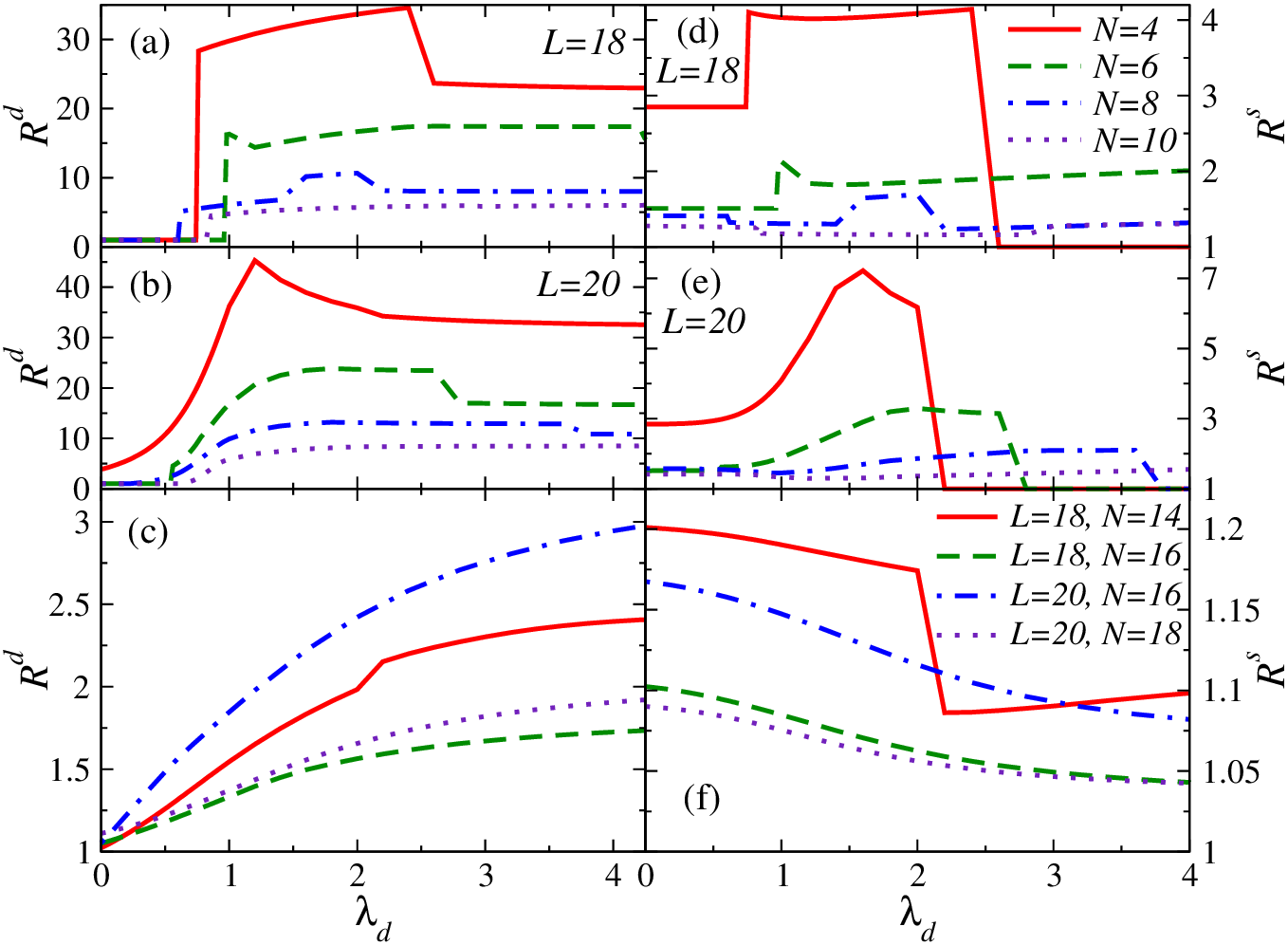}
\end{center}
\vspace{-0.6cm}
\caption{\label{fig:NaturalOrbitalsvsLambda_dwave} (Color online) 
Ratio between the two lowest natural orbitals, of the $d$-wave (left panels) 
and extended-$s$-wave (right panels) density matrices, as a function of $\lambda_d$. 
[(a) and (d)] Cluster with 18 sites and fillings $N=4,6,8,10$. [(b) and (e)] 
Cluster with 20 sites and fillings $N=4,6,8,10$. Notice that the parameter sets for 
(a)--(e) are all the same and given in (d). [(c) and (f)] Clusters with 18 and 
20 sites and two and four holes. The parameter sets in (c) and (f) are 
also the same.}
\end{figure}

Results for low electron fillings are shown in Fig.\ 
\ref{fig:FidelityvsLambda_dwave}(a) and its inset. In all cases one can 
see that there is almost no response in $g$ when $\lambda_d$ is small and 
that a strong response occurs for $\lambda_d^*\sim 1$, indicative of a 
phase transition for a finite value of $\lambda_d$. These results are 
consistent with the behavior of the ratios $R^d$ and $R^s$, which are 
depicted in Figs.\ \ref{fig:NaturalOrbitalsvsLambda_dwave}(a), 
\ref{fig:NaturalOrbitalsvsLambda_dwave}(b), \ref{fig:NaturalOrbitalsvsLambda_dwave}(d), 
and \ref{fig:NaturalOrbitalsvsLambda_dwave}(e). For most low 
fillings, both ratios change very little for small values of $\lambda_d$. 
Around $\lambda_d\sim 1$, they either jump abruptly (cluster with $L=18$) 
or increase rapidly (cluster with $L=20$). Notice that for large 
$\lambda_d$ there is almost one order of magnitude difference between the 
ratios seen for the occupation of the $d$-wave related natural orbitals 
and the extended-$s$-wave related orbitals. This is expected since the 
driving term has $d$-wave symmetry and hence $d$-wave superconductivity 
should be stabilized for large values of $\lambda_d$.

The results for low hole doping (two and four holes) are in contrast 
with those of low electron filling. Figure 
\ref{fig:FidelityvsLambda_dwave}(b) shows that in the former 
case $g$ exhibits a large response for very small values of $\lambda_d$. 
The behavior of $g$ in this case is consistent with a phase transition at 
$\lambda_d\sim 0$. The situation is similar to that of $g$ in the 
one-dimensional Hubbard model as one tunes the onsite repulsion parameter 
$U$,\cite{venuti08} where the Mott phase transition occurs at $U=0$. In 
addition, as shown in Fig.\ \ref{fig:NaturalOrbitalsvsLambda_dwave}(c) 
and \ref{fig:NaturalOrbitalsvsLambda_dwave}(f), the response of $g$ for 
small values of $\lambda_d$ is accompanied by a continuous increase of 
$R^d$ and a continuous decrease of $R^s$ for small values of $\lambda_d$. 

%%%%%%%%%%%%%%  FIGURE  %%%%%%%%%%%%%%%%%%%%%%%%%%%%%%%%%%%%%%%%%%%%%%
%%%%%%%%%%%%%%%%%%%%%%%%%%%%%%%%%%%%%%%%%%%%%%%%%%%%%%%%%%%%%%%%%%%%%%%%

Comparing the results in this subsection with Sec.\ \ref{sec:hom}, we find 
support for the view that in the thermodynamic limit, the plain $t$-$J$ model 
is superconducting (with $d$-wave symmetry, for finite values of $J$), without 
the need of introducing $\lambda_d$. Finite values of $\lambda_d$ certainly 
enhance the superconducting features of the $t$-$J$ model in finite clusters 
but may not be needed for larger system sizes. Earlier evidence in this 
direction comes from high temperature expansion studies,\cite{putikka06} and 
exact diagonalization studies of the plain $t$-$J$  
model.\cite{dagotto90,dagotto94,dagotto92,dagotto93} 

We should stress that the magnitudes of the ratios $R^d$ in
Fig.\ \ref{fig:NaturalOrbitalsvsLambda_dwave}(c), 
reveals a very important characteristic of the $d$-wave superconducting 
state generated in a lightly doped Mott insulator. The $d$-wave 
condensate occupation is low if one compares it with the one generated, 
for low electron filling [Figs.\ \ref{fig:NaturalOrbitalsvsLambda_dwave}(a) 
and  \ref{fig:NaturalOrbitalsvsLambda_dwave}(b)], under the influence 
of a suitable value of $\lambda_d$. This low condensate 
occupation is a direct consequence of the strong correlations present in 
the system and is similar to the behavior of the condensate occupation in 
liquid helium, which is known to be strongly depleted due to the effects 
of strong interactions. 

\subsection{$s$-wave term}

\begin{figure}[!h]
\begin{center}
  \includegraphics[width=0.41\textwidth,angle=0]{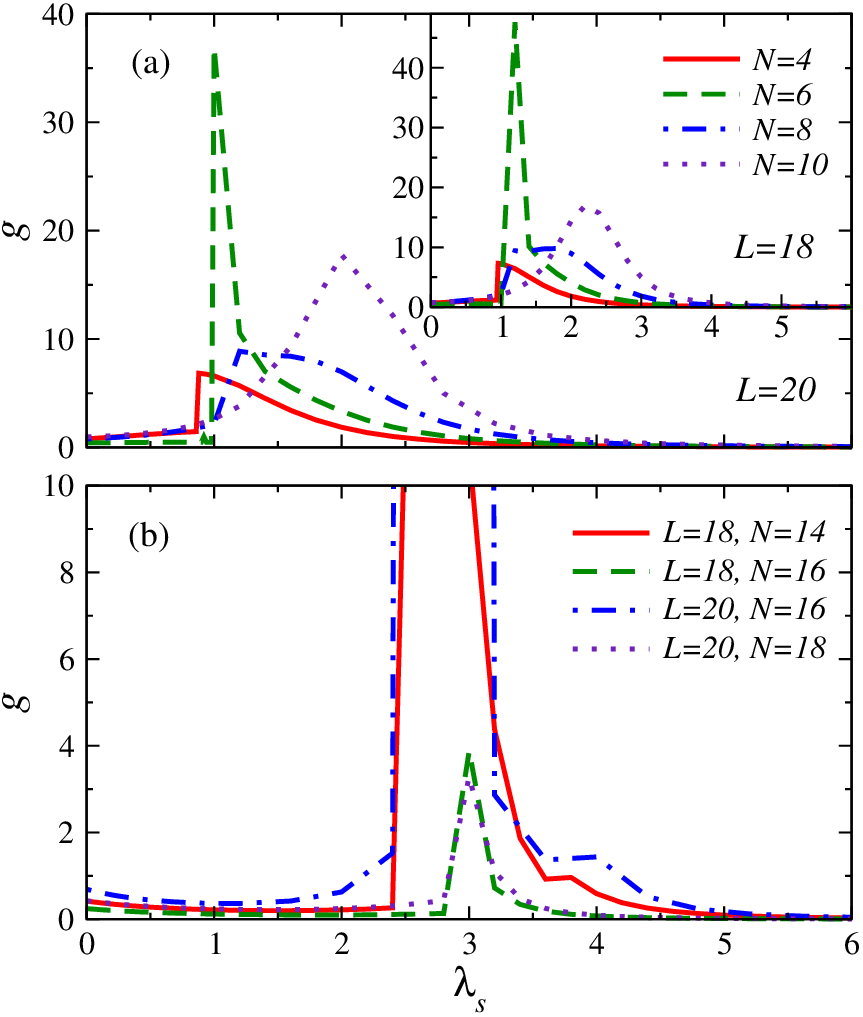}
\end{center}
\vspace{-0.6cm}
\caption{\label{fig:FidelityvsLambda_swave} (Color online) Fidelity 
metric of the $t$-$J$ model ($J=0.3$) after the addition of an 
extended-$s$-wave-superconductivity inducing term. (a) Cluster with 20 
sites and fillings $N=4,6,8,10$. (b) Clusters with 18 and 20 sites and 
two and four holes. The inset in (a) depicts the results for the cluster 
with 18 sites and fillings $N=4,6,8,10$.}
\end{figure}

In order to further study the suggestions in Sec.\ \ref{sec:supd}, we 
have also studied the effect of adding an  
extended-$s$-wave-superconductivity inducing term 
[Eq.\ (\ref{eq:dwave})] to the plain $t$-$J$ model. In Fig.\ 
\ref{fig:FidelityvsLambda_swave}, 
we explore the behavior of the fidelity metric for such a model.

%%%%%%%%%%%%%%  FIGURE  %%%%%%%%%%%%%%%%%%%%%%%%%%%%%%%%%%%%%%%%%%%%%%
%%%%%%%%%%%%%%%%%%%%%%%%%%%%%%%%%%%%%%%%%%%%%%%%%%%%%%%%%%%%%%%%%%%%%%%%

\begin{figure}[!h]
\begin{center}
  \includegraphics[width=0.48\textwidth,angle=0]{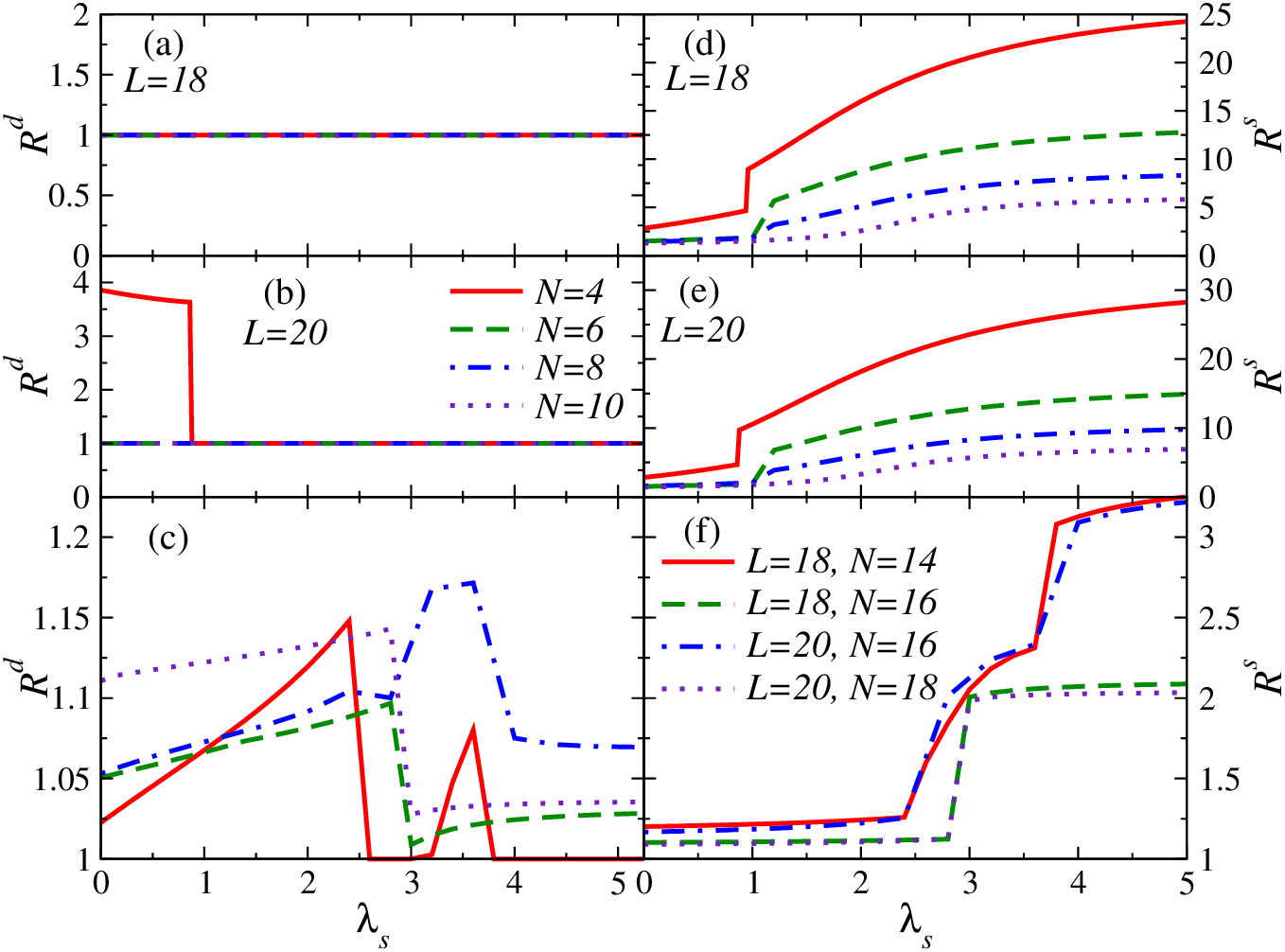}
\end{center}
\vspace{-0.6cm}
\caption{\label{fig:NaturalOrbitalsvsLambda_swave} (Color online) 
Ratio between the two lowest natural orbitals, of the $d$-wave (left panels) 
and extended-$s$-wave (right panels) density matrices, as a function of 
$\lambda_s$. [(a) and (d)] Cluster with 18 sites and fillings $N=4,6,8,10$.
[(b) and (e)] Cluster with 20 sites and fillings $N=4,6,8,10$. Notice that the 
parameter sets for (a)--(e) are all the same and given in (d). [(c) and (f)] 
Clusters with 18 and 20 sites and two and four holes. The parameter sets 
in (c) and (f) are also the same.}
\end{figure}

Figure \ref{fig:FidelityvsLambda_swave} shows that in this case small 
values of the control parameter $\lambda_s \leq 1$ induce almost no response in 
the fidelity metric at any filling. Instead, one needs a sizable magnitude 
of $\lambda_s$, at both low electron filling and low hole doping, to 
trigger a large response in $g$, which would signal a phase transition 
to an extended-$s$-wave superconducting state. The calculated ratio 
between the two lowest natural orbitals, depicted in 
Fig.\ \ref{fig:NaturalOrbitalsvsLambda_swave}, is also consistent with 
this observation. Only large values of $\lambda_s$ enhance the 
extended-$s$-wave condensate occupation for all fillings [Figs.\ 
\ref{fig:NaturalOrbitalsvsLambda_swave}(d)--\ref{fig:NaturalOrbitalsvsLambda_swave}(f)],
particularly for low hole doping [Fig.\ \ref{fig:NaturalOrbitalsvsLambda_swave}(f)]. 
Interestingly, in analogy to the $d$-wave case for $R^d$, the enhancement of the ratio 
$R^s$ is almost an order of magnitude larger at low electron fillings 
than at low hole doping, evidencing the strong depletion of the condensate 
that occurs in the latter case due to the presence of strong correlations. 

%%%%%%%%%%%%%%  FIGURE  %%%%%%%%%%%%%%%%%%%%%%%%%%%%%%%%%%%%%%%%%%%%%%
%%%%%%%%%%%%%%%%%%%%%%%%%%%%%%%%%%%%%%%%%%%%%%%%%%%%%%%%%%%%%%%%%%%%%%%%

Overall, these findings are consistent with previous exact diagonalization 
studies of the $t$-$J$ model\cite{dagotto90,dagotto92,dagotto93} that 
have provided numerical evidence for dominant $d$-wave pairing at low 
hole doping, absent in the opposite limit of low electron filling. Our 
results also show that strong correlations become more important as one 
approaches half-filling, where no matter which superconductivity 
enhancing term one introduces, the resulting condensate occupation is 
always small.

\section{Quenched Disorder}
\label{sec:dis}

In this section, we address the question of the effect of disorder in the 
models and quantities studied in the previous sections for the clean case.

\begin{figure}[!ht]
\begin{center}
  \includegraphics[width=0.41\textwidth,angle=0]{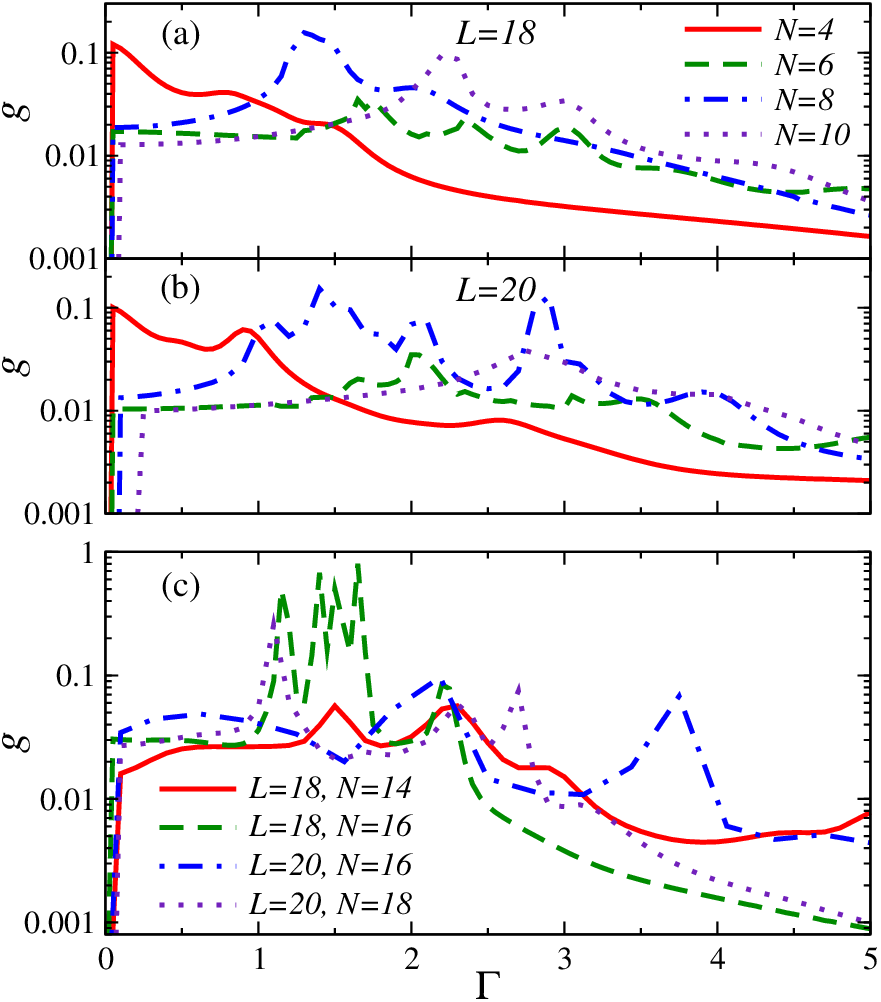}
\end{center}
\vspace{-0.6cm}
\caption{\label{fig:FidelityvsDisorder_tJ} (Color online) Fidelity metric 
of the $t$-$J$ model ($J=0.3$) as a function of the disorder strength 
$\Gamma$. (a) Cluster with 18 sites and fillings $N=4,6,8,10$. 
(b) Cluster with 20 sites and fillings $N=4,6,8,10$ [the same parameter 
sets as in (a)]. (c) Clusters with 18 and 20 sites and two and four 
holes. In all cases the fidelity metric was computed averaging over 20 
different disorder realizations, except for the cluster with $L=20$ and 
fillings $N=10$ and $N=16$ where the average was performed over ten disorder 
realizations.}
\end{figure}

For the plain $t$-$J$ model, in Sec.\ \ref{sec:hom}, we have seen that 
the lowest natural orbital and the ratio between the lowest and second 
lowest natural orbitals are in general small. For that model, we focus 
here on the response of the fidelity metric to adding disorder to the 
system. Results for this quantity, averaged over different disorder 
realizations, are presented in Fig.\ \ref{fig:FidelityvsDisorder_tJ}.
For weak interactions, one would expect small values of $\Gamma$ to 
produce localization and dramatically change the nature of the ground 
state in the clean case. In contrast, for strong correlations,
this figure illustrates that, except for the very low electron-filling 
case of four particles, one needs a very large disorder strength 
($\Gamma>1$) in order to trigger a large response in $g$. Different 
disorder realizations produce a large response in $g$ for different 
values of $\Gamma$, hence, the various peaks that can be seen in many of 
the plots (notice that the results for $g$ are presented in a semi-log 
scale). However, the common feature for all fillings (with the exception 
of $N=4$) is that large responses only occur when $\Gamma>1$. This 
indicates that the nature of the ground state of the system in these
cases is robust against nonmagnetic disorder and 
exemplifies the importance of strong correlations in the system. 
The exception is the case of very low fillings [e.g., 
$N=4$ for $L=18$ and $L=20$ in Fig.\ \ref{fig:FidelityvsDisorder_tJ}(a) 
and \ref{fig:FidelityvsDisorder_tJ}(b), respectively] in the $t$-$J$ 
model, where the Gutzwiller projection plays a subdominant role.

%%%%%%%%%%%%%%  FIGURE  %%%%%%%%%%%%%%%%%%%%%%%%%%%%%%%%%%%%%%%%%%%%%%
%%%%%%%%%%%%%%%%%%%%%%%%%%%%%%%%%%%%%%%%%%%%%%%%%%%%%%%%%%%%%%%%%%%%%%%%

In the clean case, the addition of a $d$-wave superconductivity inducing 
term was used to precipitate $d$-wave superconductivity in the $t$-$J$ 
model and was shown to produce a large enhancement of the ratio $R^d$ for 
low electron fillings and a smaller enhancement for low hole doping. Here, 
we study the effect of disorder in both the fidelity metric $g$ and the 
ratio $R^d$ when $J=0.3$ and $\lambda_d=1.25$.

\begin{figure}[!ht]
\begin{center}
  \includegraphics[width=0.48\textwidth,angle=0]{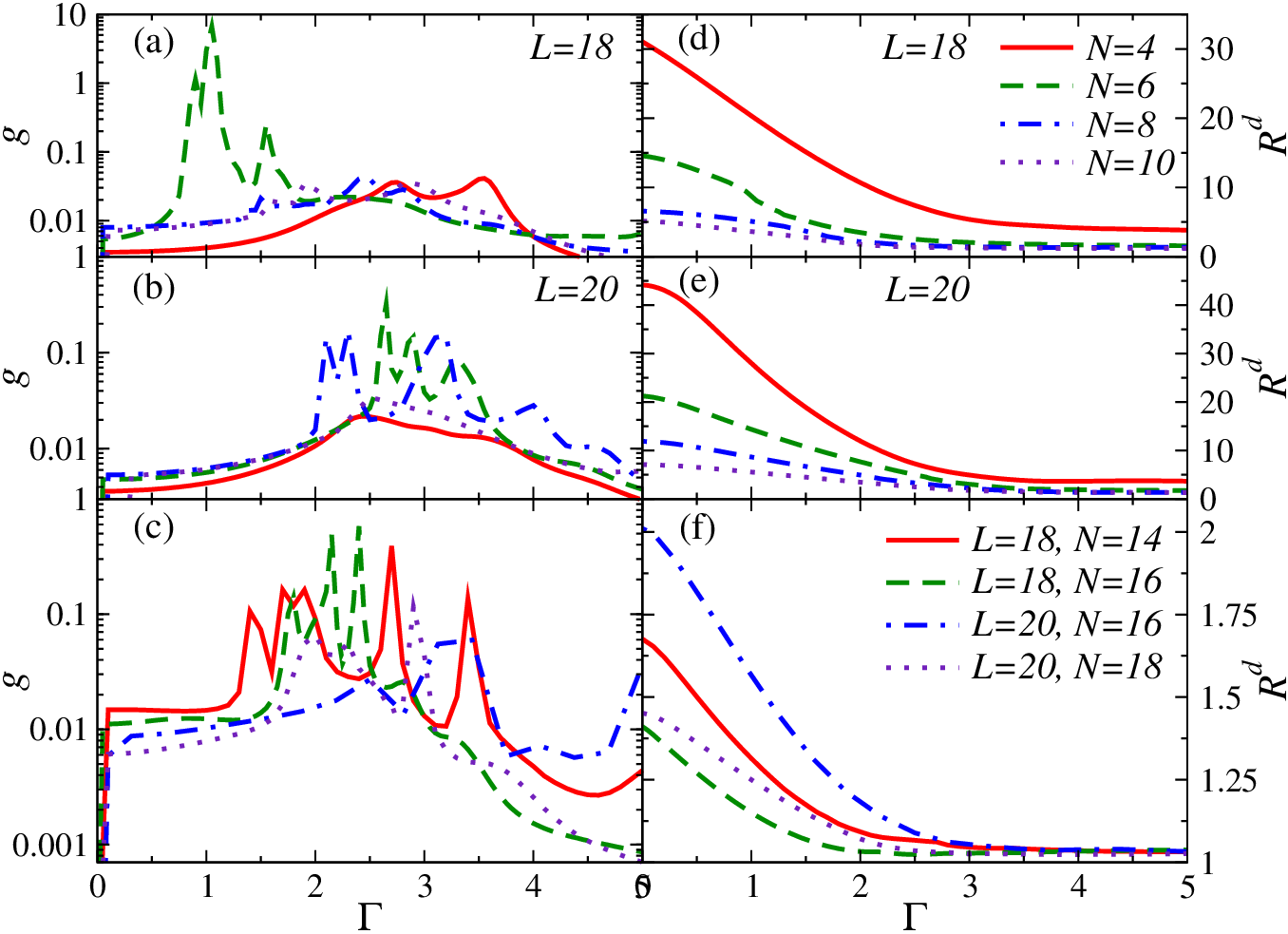}
\end{center}
\vspace{-0.6cm}
\caption{\label{fig:FidelityvsDisorder_tJdwave} (Color online) Fidelity 
metric (left column) and ratio between the two largest eigenvalues of the 
$d$-wave density matrix (right column) for the $t$-$J$ model ($J=0.3$) 
with the addition of a $d$-wave superconductivity inducing term 
($\lambda_d=1.25$) as a function of the disorder strength $\Gamma$. 
[(a) and (d)] Cluster with 18 sites and fillings $N=4,6,8,10$. [(b) and (e)] 
Cluster with 20 sites and fillings $N=4,6,8,10$ [the same parameter sets as in 
(a) and (d)]. [(c) and (f)] Clusters with 18 and 20 sites and two and four holes.
In all cases $g$ and $R^d$ ware computed averaging over 20 
different disorder realizations, except for the cluster with $L=20$ and 
fillings $N=10$ and $N=16$ where the average was performed over ten disorder 
realizations.}
\end{figure}

In Fig.\ \ref{fig:FidelityvsDisorder_tJdwave}, we have plotted the 
evolution of $g$ (left panel) and $R^d$ (right panel) with increasing 
disorder. The results depicted in this figure were obtained averaging 
over the same number of disorder realizations as in Fig.\ 
\ref{fig:FidelityvsDisorder_tJ}. By comparing $g$ in Figs.\ 
\ref{fig:FidelityvsDisorder_tJ} and \ref{fig:FidelityvsDisorder_tJdwave}, 
one can see that the effect of adding a finite value of $\lambda_d=1.25$ 
is more pronounced at low fillings, where large responses in $g$ move 
toward larger values of $\Gamma$ ($\Gamma>2$, with the exception of $N=6$ 
in the $L=18$ cluster that still has a peak around $\Gamma=1$). These 
large responses are expected to indicate of the destruction of the superconducting 
ground state generated in the clean case. This can be better seen in Figs.\ 
\ref{fig:FidelityvsDisorder_tJdwave}(d) and 
\ref{fig:FidelityvsDisorder_tJdwave}(e), where the value of $R^d$ is much 
smaller than the one in the clean case whenever a large response is seen 
in $g$ [Figs.\ \ref{fig:FidelityvsDisorder_tJdwave}(a) and 
\ref{fig:FidelityvsDisorder_tJdwave}(b)].

Interestingly, in the low hole-doping regime, the behavior of the fidelity 
metric as a function of disorder is very similar when $\lambda_d=1.25$ 
and in the clean case, an indication that in this regime the ground state 
of the $t$-$J$ model is not dramatically affected by the presence of 
$\lambda_d$. This marked contrast with the low electron filling is 
further supported if one realizes that the relative reduction of $R^d$ is 
much more drastic at low electron filling than in the low hole-doping 
regime. Further studies using alternative numerical approaches may be 
needed in to clarify the scaling of these effects with system size.

Overall, our finding in the presence or absence of the superconductivity 
inducing term is that strong correlations generate a ground state in 
the $t$-$J$ model that is robust against disorder.

%%%%%%%%%%%%%%  FIGURE  %%%%%%%%%%%%%%%%%%%%%%%%%%%%%%%%%%%%%%%%%%%%%%
%%%%%%%%%%%%%%%%%%%%%%%%%%%%%%%%%%%%%%%%%%%%%%%%%%%%%%%%%%%%%%%%%%%%%%%%

\section{Summary}
\label{sec:sum}

Within the $t$-$J$ model, we have calculated the ground-state fidelity, 
the $d$-wave and extended-$s$-wave condensate occupations, with and 
without the addition of a superconductivity inducing term, and in the 
clean and disordered cases.

In the clean case we find that: 

(i) the plain $t$-$J$ model exhibits a distinctive signature in the 
fidelity metric at low hole doping and small values of $J$ that may be an 
indication that a continuous phase transition occurs in this regime. By 
studying the $d$-wave and extended-$s$-wave condensate occupations, we 
find that the former is favored. As expected, spin-spin 
(antiferromagnetic) correlations are enhanced by $J$ in the low 
hole-doping regime, but they are drastically reduced by doping. 
Density-density correlations, on the other hand, do not exhibit any clear 
signature of order as $J$ is increased. Adding a $t'<0$ term only moves 
the response of the fidelity metric towards larger values of $J$. 
This is indicative of a continuous phase transition for a value of $J$ 
that is larger than for the plain $t$-$J$ model.

(ii) In order to better understand the nature of the phases in the plain 
$t$-$J$ model, we added infinite range $d$-wave and extended-$s$-wave 
superconductivity inducing terms. For a finite value of $J=0.3$, we find 
that at low-hole doping arbitrarily small $d$-wave driving terms induce a 
large response in the fidelity metric, which is consistent with the plain 
$t$-$J$ model having a $d$-wave superconducting ground state. In 
contrast, almost no response was found for low electron filling, and 
for the extended-$s$-wave superconductivity driving term for all 
fillings. In these latter cases one needs a large value of the driving term 
in order to obtain a sizable response. Interestingly, at low hole doping, 
we always find the condensate occupations to be small, i.e., there is a 
strong depletion of the (relevant) condensate no matter the symmetry of 
the driving term. This is an indication that, for the plain $t$-$J$ model, 
mean-field theories based on the $d$-wave order parameter may not be 
justified.

In the dirty case, we have shown that very strong disorder is required to 
produce large changes in the ground state of the plain $t$-$J$ model 
whenever the electron filling  is not very low, i.e., whenever the system 
is strongly correlated. Adding a $d$-wave superconductivity inducing term 
was shown to leave the results for the low hole-doping regime almost 
unchanged with respect to the ones of the plain $t$-$J$ model, while 
the results for the low electron-filling regime were modified to a larger 
extent. This is, once again, consistent with the hypothesis that the plain 
$t$-$J$ model has a $d$-wave superconducting ground state for low hole 
doping. Future studies on the scaling of the effects and observables 
discussed here, with system size, could shed further light on the nature 
of the ground state of the $t$-$J$ model and its relation to 
high-temperature superconductivity.

\begin{acknowledgments}

We acknowledge support from a Startup Fund (M.R. at Georgetown University),  
DOE-BES Grant No.\ DE-FG02-06ER46319 (B.S.S. at  UCSC), 
and NSF Grant No.\ DMR-0804914 (S.H. at USC). We thank 
G. H. Gweon, A. Muramatsu, M. Randeria, J. A. Riera, and R. T. Scalettar 
for helpful discussions. Computational facilities were provided by the High 
Performance Computing and Communications Center at the University of 
Southern California. M.R. is grateful to the Aspen Center for Physics 
where this work was finalized.
\end{acknowledgments}

\begin{figure}[!h]
\begin{center}
  \includegraphics[width=0.48\textwidth,angle=0]{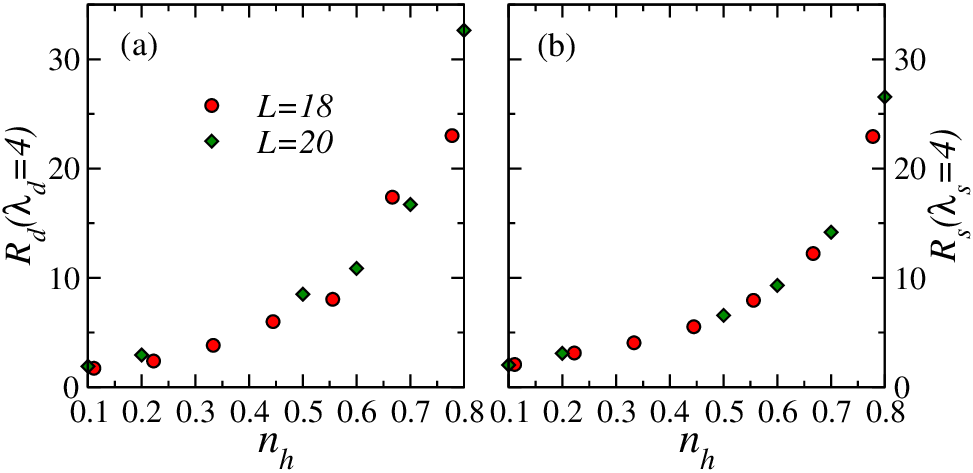}
\end{center}
\vspace{-0.6cm}
\caption{\label{fig:NaturalOrbitalsvsnh} (Color online) 
Ratio between the two lowest natural orbitals, of the $d$-wave (a) 
and extended-$s$-wave (b) density matrices, as a function of hole doping 
($n_h$) for a large superconductivity inducing term with the same 
symmetry of the order parameter plotted. Results are presented for clusters 
with 18 and 20 sites. For low hole doping one can see a nearly linear 
dependence of the ratios $R^d$ and $R^s$ with $n_h$.}
\end{figure}

{\it Note added on proof.} --With regard to the ratios (condensate occupations) 
$R^d$ and $R^s$ in the low hole doping regime, we would like 
to emphasize that in addition to being small (even in the presence of large 
superconductivity enhancing terms in the Hamiltonian) they increase linearly 
with increasing doping. This is shown in Fig.\ \ref{fig:NaturalOrbitalsvsnh}.

%%%%%%%%%%%%%%  FIGURE  %%%%%%%%%%%%%%%%%%%%%%%%%%%%%%%%%%%%%%%%%%%%%%
%%%%%%%%%%%%%%%%%%%%%%%%%%%%%%%%%%%%%%%%%%%%%%%%%%%%%%%%%%%%%%%%%%%%%%%%


\begin{thebibliography}{99}
\bibitem{notes}
We should keep in mind that for finite size systems true long ranged order 
is impossible. In fact, the superconducting BCS type state projected to a 
fixed number would have a non-vanishing overlap with the normal state, 
especially for small system sizes.
\end{thebibliography}
\end{document}